\documentstyle[12pt]{article}
\topmargin - 0.8cm

\renewcommand{\theequation}{\arabic{section}.\arabic{equation}}
\renewcommand{\thesection}{\arabic{section}.}

\catcode`@=11
\@addtoreset{equation}{section}
\catcode`@=12
\begin{document}
\title{\vskip-1.7cm \bf Tunnelling geometries II. Reduction methods
for
functional determinants}
\author{A. O. Barvinsky\thanks{On leave from Nuclear Safety
Institute, Russian Academy of Sciences , Bolshaya Tulskaya 52, Moscow
113191, Russia}}
\date{}
\maketitle
{\em
Theoretical Physics Institute, Department of Physics, \ University of
Alberta, Edmonton, Canada T6G 2J1}
Moscow
\begin{abstract}
The reduction algorithms for functional determinants of differential
operators
on spacetime manifolds of different topological types are presented,
which were
recently used for the calculation of the no-boundary wavefunction and
the
partition function of tunnelling geometries in quantum gravity and
cosmology.
\end{abstract}

\section{Introduction}
\hspace{\parindent}
In this paper we present the calculational technique for functional
determinants of differential operators arising in the one-loop
approximation of
local field theory on curved spacetime manifolds of different
topological
types. The purpose of this technique consists in the reduction method
which
allows one to obtain these determinants in terms of quantities of
lower
functional dimensionality, that is the determinants and traces on the
space of
functions defined at some spatial slice and not involving the
integration over
time. This method, well known in the context of the semiclassical WKB
theory as
a Pauli-Van Vleck-Morette formula \cite{Pauli}, has a rather
nontrivial
generalization to Hartle-Hawking boundary conditions describing the
quantum
tunnelling geometries in quantum cosmology. It was applied in
\cite{BKam:norm,tunnelI} for the calculation of the no-boundary
wavefunction
and served as a link between the unitary Lorentzian theory and the
gravitational instantons in the Euclidean quantum gravity.

The organization of the paper is as follows. In Sect.2 we give a
short overview
of the functional determinants arising in the one-loop approximation
for the
path integral which gives the transition amplitude for different
types of
boundary conditions. Sect.3 contains the derivation of the known
Pauli-Van
Vleck-Morette formula for a one-loop preexponential factor,
appropriate for its
generalization to the case of the no-boundary wavefunction in quantum
cosmology. In Sect.3 we obtain the reduction algorithm for its
preexponential
factor in terms of the regular basis functions of the Euclidean
"wave" operator
on the underlying spacetime of the no-boundary type. In Sect.4 we
derive a
similar reduction algorithm for the preexponential factor on the
closed
Euclidean spacetime without boundary, having a topology of a
four-dimensional
sphere. We also discuss a deep analogy of this result with the known
algorithm
for a transition amplitude between IN and OUT vacua of the S-matrix
theory in
the asymptotically flat Lorentzian spacetime, which underlies the
unification
of the unitary Lorentzian quantum gravity with its Euclidean version
considered
in \cite{tunnelI}.

\section{Functional determinants in the one-loop approximation of
quantum
theory}
\hspace{\parindent}
The one-loop functional determinants in Euclidean quantum theory, and
quantum
gravity in particular, arise in context of the heat equation
	\begin{eqnarray}
	&&\hbar\,\frac{\partial}{\partial\tau_{+}}
	K(\tau_{+},\phi_{+}|\,\tau_{-},\phi_{-}\!)=-\hat H\,
	K(\tau_{+},\phi_{+}|\,\tau_{-},\phi_{-}\!),
\label{eqn:6.29}\\
	&&K(\tau_{-},\phi_{+}|\,\tau_{-},\phi_{-}\!)=
	\delta\,(\phi_{+}-\phi_{-}\!)
\label{eqn:6.30}
	\end{eqnarray}
for a kernel of transition between the configurations $\phi_{\pm}$ at
the
"moments" of the Euclidean time $\tau_{\pm}$, generated by the
Hamiltonian of
the theory $\hat H=\hat H\,(\phi_{+},p_{\phi})$ which is a
differential
operator in the coordinate representation of $\phi_{+}$,
$p_{\phi}=\hbar\partial/i\partial\phi_{+}$. This kernel can be
represented by
the path integral over histories $\phi\,(\tau)$ subject to boundary
conditions
$\phi\,(\tau_{\pm})=\phi_{\pm}$
      \begin{equation}
      K(\tau_{+},\phi_{+}|\,\tau_{-},\phi_{-}\!)=
      \int D\mu[\,\phi\,]\,\,
      {\rm e}^{\!\!\phantom{0}^{\textstyle
      -\frac{1}{\hbar}I\,[\,\phi\,]}}
\label{eqn:2.2}
      \end{equation}
with the Euclidean action
	\begin{eqnarray}
	 I\,[\,\phi\,(\tau)\,]=
	 \int\limits_{\tau_{-}}^{\tau_{+}}d\tau\,
	 {\cal{L}}_{E}(\phi,d\phi/d\tau,\tau)
\label{eqn:3.11}
	 \end{eqnarray}
and local integration measure
 	\begin{eqnarray}
	&&D\mu[\,\phi\,]=\prod_{\tau}d\phi\,(\tau)\;
	[\,{\rm det}\,a\,]^{1/2}\!(\tau)
	+O\,(\,\hbar\,),\;\;\;
	d\phi=\prod_{i}d\phi^{i},
\label{eqn:3.2}\\
	&&{\rm det}\,a={\rm det}\,a_{ik}, \;\;\;
	a_{ik}=\frac{\partial^{2}{\cal L}_{E}}
	{\partial\dot \phi^{i}\partial\dot \phi^{k}},
\label{eqn:3.4}
	\end{eqnarray}
where the Euclidean Lagrangian is related to the Hamiltonian $
H\,(\phi_{+},p_{\phi})$ by the Legendre transform with respect to
$ip_{\phi}$
\footnote
{The equations (\ref{eqn:6.29}) - (\ref{eqn:3.4}) are related to the
physical
theory in Lorentzian time by the Wick rotation of time variable
$\tau=it+{\rm
const}$ accounting, in particular, for the occurence of imaginary
unit in the
transform from the Hamiltonian to the Euclidean Lagrangian.
}.

In theories with the Hamiltonian and Lagrangian quadratic
respectively in
momenta and velocities, the integration measure (\ref{eqn:3.2}) does
not
include the higher-order loop corrections $O\,(\hbar)$ and reduces to
the
determinant of the Hessian matrix $a_{ik}$. The latter is understood
with
respect to indices $i$ and $k$ labelling the physical variables
$\phi=\phi^{i}$. Throughout the paper we shall use DeWitt notations,
in which
these indices have a condensed nature and include, depending on the
representation of field variables, either continuous labels of
spatial
coordinates or discrete quantum numbers labelling some complete
infinite set of
harmonics on a spatial section of spacetime. Correspondingly the
contraction of
these indices, the trace operation, ${\rm tr}\,A\equiv A^{i}_{i}$,
with respect
to them, etc. will imply either the spatial integration or infinite
summation
over such a set. Therefore the above determinant in field systems is
functional, but its functional nature is restricted to a spatial
slice of
constant time $\tau$. The product over time points of ${\rm
det}\,a(\tau)$ can
be regarded as a determinant of higher functional dimensionality
associated
with the whole spacetime if we redefine $a_{ik}$ as a time-ultralocal
operator
${\mbox{\boldmath $a$}}= a_{ik}\delta\,(\tau-\tau')$. We shall denote
such
functional determinants for both ultralocal and differential
operators in time
by Det. Thus, in view of the ultralocality of ${\mbox{\boldmath
$a$}}$, the
contribution of the one-loop measure equals
        \begin{equation}
	\prod_{t}\,[\,{\rm det}\,a\,]^{1/2}(t)=
	[\,{\rm Det}\,{\mbox{\boldmath $a$}}\,]^{1/2}=
	{\rm exp}\left \{\frac{1}{2}
	\int_{t_{-}}^{t_{+}}\!dt\,
	\delta\,(0)\;{\rm ln\,det}\,a(t)\right\}.
\label{eqn:3.6}
	\end{equation}

With these notations and local measure, the one-loop approximation
for the path
inegral (\ref{eqn:2.2}) takes the form
	\begin{equation}
	 K_{\rm 1-loop}(\tau_{+},\phi_{+}|\,\tau_{-},\phi_{-}\!)=
	\left(\,\frac{{\rm Det}\,
	{\mbox{\boldmath $F$}}}{{\rm
	 Det}\,{\mbox{\boldmath $a$}}}\,\right)^{-1/2}
	 \left.{\rm e}^{\!\!\phantom{0}^{\textstyle
         -\frac{1}{\hbar}I\,[\,\phi\,]}}\,
	 \right|_
	 {\;\phi\,(\tau)=\phi\,(\tau,\,\phi_{\pm}\!)},
\label{eqn:3.18}
	 \end{equation}
where $\phi\,(\tau,\,\phi_{\pm}\!)$ is a solution of classical
equations of
motion passing the points $\phi_{\pm}$ at $\tau_{\pm}$
	\begin{equation}
	\frac{\delta I\,[\,\phi\,]}{\delta\phi\,(\tau)}=0,\;\;\;
	\phi\,(\tau_{-},\,\phi_{\pm}\!)=\phi_{-},\;\;\;
	\phi\,(\tau_{+},\,\phi_{\pm}\!)=\phi_{+}
\label{eqn:3.18a}
	\end{equation}
and ${\mbox{\boldmath $F$}}$ is an operator of small disturbances on
its
background -- a matrix-valued second-order differential operator
${\mbox{\boldmath $F$}}\,(d/d\tau)\equiv {\mbox{\boldmath
$F$}}_{ik}(d/d\tau)$
       \begin{equation}
       {\mbox{\boldmath $F$}}\equiv{\mbox{\boldmath
$F$}}\,(d/d\tau\!)\,
       \delta(\tau-\tau^{\prime})=
       \frac{\delta^{2}I\,[\,\phi\,]}
       {\delta\phi(\tau)\,
       \delta\phi(\tau^{\prime})}.                   \label{eqn:3.21}
       \end{equation}
Generically it has the form
	\begin{equation}
       {\mbox{\boldmath $F$}}\,(d/d\tau)=-\frac{d}{d\tau}\,
       a\,\frac{d}{d\tau}
       -\frac{d}{d\tau}\,b+b^{T}\frac{d}{d\tau}+c,
\label{eqn:3.22}
       \end{equation}
where the coefficients $a=a\,_{ik},\;b=b\,_{ik}$ and $c=c\,_{ik}$ are
the
(functional) matrices acting in the space of field variables
$\phi\,(\tau)=\phi\,^{k}(\tau)$, and the superscript $T$ denotes
their
(functional) transposition $(b^{T})\,_{ik}\equiv b\,_{ki}$. These
coefficients
can be easily expressed as mixed second-order derivatives of the
Euclidean
Lagrangian with respect to $\phi\,^{i}$ and
$\dot\phi\,^{i}=d\phi\,^{i}/d\tau$.
In particular, the matrix of the second order derivatives $a\,_{ik}$
is given
by the Hessian matrix (\ref{eqn:3.4}): $a\,_{ik}=\partial^{2}{\cal
L}_{E}/\partial\dot\phi\,^{i}\partial\dot\phi\, ^{k}$.

In the general case, when the spectrum of a differential operator is
unknown,
the only definition for its functional determinant can be given via
the
following variational equation
         \begin{equation}
	 \delta\,{\rm ln}\,{\rm Det}\,{\mbox{\boldmath $F$}}=
	 \delta\,{\rm Tr}\,{\rm ln}\,{\mbox{\boldmath $F$}}=
	 {\rm Tr}\,
	 \delta{\mbox{\boldmath $F$}}\,{\mbox{\boldmath $G$}},
\label{eqn:6.1}
	 \end{equation}
where ${\rm Tr}$ denotes the functional trace, which includes the
integration
over time of the coincidence limit of the corresponding operator
kernel, and
${\mbox{\boldmath $G$}}={\mbox{\boldmath
$G$}}\,(\tau,\tau^{\prime}\!)$ is an
inverse of ${\mbox{\boldmath $F$}}$ or the Green's function
satisfying
          \begin{equation}
	  {\mbox{\boldmath $F$}}(d/d\tau\!)\,{\mbox{\boldmath
$G$}}\,(\tau,\tau^{\prime}\!)=
	  \delta\,(\tau,\tau^{\prime}\!).
\label{eqn:6.2}
	  \end{equation}
This definition is obviously incomplete unless one fixes uniquely the
Green's
function by the appropriate boundary conditions and also specifies
the
functional composition law $\delta{\mbox{\boldmath
$F$}}\,{\mbox{\boldmath
$G$}}$ in the functional trace. One should remember that the kernel
${\mbox{\boldmath $G$}}\,(\tau,\tau^{\prime}\!)$ is not a smooth
function of
its arguments, and its irregularity enhances  when it is acted upon
by two
derivatives contained in $\delta{\mbox{\boldmath
$F$}}=\delta{\mbox{\boldmath
$F$}} (d/d\tau\!)$, therefore one has to prescribe the way how these
derivatives act on both arguments of  ${\mbox{\boldmath
$G$}}\,(\tau,\tau^{\prime}\!)$ and how to take the coincidence limit
of the
resulting singular kernel in the functional trace.

Both the boundary conditions and the specification of trace operation
must
follow from the way of calculating the gaussian path integral over
quantum
disturbances $\varphi\,(\tau)$ in the vicinity of
$\phi\,(\tau)=\phi\,(\tau,\,\phi_{\pm}\!)$, which gives rise to the
one-loop
functional determinants of the above type:
          \begin{equation}
	  \int D\varphi\,
           \,{\rm exp}
          \left\{-\frac{1}{2\hbar}\int_{\tau_{-}}
	  ^{\tau_{+}}d\tau\,\varphi^{T}\!\stackrel
	  {\leftrightarrow}{{\mbox{\boldmath $F$}}}\!\varphi\right\}
	  ={\rm Const}\,\left[\,{\rm Det}\,{\mbox{\boldmath $F$}}\,
	  \right]^{-1/2}.
\label{eqn:6.3}
           \end{equation}
Here the notation $\stackrel{\leftrightarrow}{{\mbox{\boldmath
$F$}}}$ implies
that the two time derivatives of the operator ${\mbox{\boldmath
$F$}}$ are
acting symmetrically on the both functions $\varphi$ and
$\varphi^{T}$, so that
the exponentiated time integral contains the quadratic part of the
Euclidean
Lagrangian ${\cal L}_{E}^{(2)}=(1/2)\,\varphi^{T}
\!\stackrel{\leftrightarrow}
{{\mbox{\boldmath $F$}}}\!\varphi$, generating the "wave" operator
${\mbox{\boldmath $F$}}$. For arbitrary two test functions
$\varphi_{1}$ and
$\varphi_{2}$ and for the operator of the form (\ref{eqn:3.22}) this
notation
reads
           \begin{equation}
	   \varphi^{T}_{1}\!\stackrel
	   {\leftrightarrow}{{\mbox{\boldmath $F$}}}\!\varphi_{2}=
	   \dot\varphi^{T}_{1}\,a\,\dot\varphi_{2}+
	   \dot\varphi^{T}_{1}\,b\,\varphi_{2}+
	   \varphi^{T}_{1}\,b^{T}\dot\varphi_{2}+
	   \varphi^{T}_{1}c\,\varphi_{2}
\label{eqn:6.4}
	   \end{equation}
and implies the following integration by parts
           \begin{eqnarray}
	   &&\varphi^{T}_{1}\!\stackrel
	   {\leftrightarrow}{{\mbox{\boldmath $F$}}}\!\varphi_{2}=
	   \varphi^{T}_{1}\,({{\mbox{\boldmath $F$}}}\varphi_{2})+
	   \frac{d}{d\tau}\left[\,\varphi^{T}_{1}\,
	   ({{\mbox{\boldmath $W$}}}\!\varphi_{2})\,\right],
\label{eqn:6.5}\\
	   &&{\mbox{\boldmath $W$}}\equiv{\mbox{\boldmath
$W$}}(d/d\tau)
	   =a\,\frac{d}{d\tau}+b.
\label{eqn:6.5a}
	   \end{eqnarray}
We shall call ${\mbox{\boldmath $W$}}$ the {\it Wronskian} operator
which
enters the following Wronskian relation for the operator
${\mbox{\boldmath
$F$}}$
          \begin{equation}
	  \varphi^{T}_{1}\,({\mbox{\boldmath $F$}}\varphi_{2}\!)-
	  ({\mbox{\boldmath $F$}}\varphi_{1}\!)^{T}\varphi_{2}=
	  -\frac{d}{d\tau}\left[\,\varphi^{T}_{1}\,
	  ({\mbox{\boldmath $W$}}\varphi_{2}\!)-({\mbox{\boldmath
$W$}}\varphi_{1}\!)^{T}
	  \varphi_{2}\,\right]
\label{eqn:6.6}
	  \end{equation}
and also participates in the variational equation for the Euclidean
momentum
$\partial{\cal{L}}_{E}/\partial\dot\phi$ valid for arbitrary
variations
$\delta\phi\,(\tau)$ of field histories
	  \begin{eqnarray}
	  \delta\frac{\partial{\cal{L}}_{E}}
          {\partial\dot\phi}
          ={\mbox{\boldmath $W$}}(d/d\tau\!)\,\delta\phi\,(\tau).
\label{eqn:4.8}
          \end{eqnarray}

As shown by Feynman \cite{Feynman}, the functional determinant of the
differential operator generated by the gaussian path integral
(\ref{eqn:6.3})
is determined by the variational equation
         \begin{equation}
	 \delta\,{\rm ln}\,{\rm Det}\,{\mbox{\boldmath $F$}}=
	 {\rm Tr}\,\stackrel{\leftrightarrow}
	 {\delta{\mbox{\boldmath $F$}}}\!{\mbox{\boldmath $G$}}
\label{eqn:6.7}
	 \end{equation}
where the Green's function ${\mbox{\boldmath
$G$}}\,(\tau,\tau^{\prime}\!)$
satisfies the {\it same} boundary conditions at $\tau=\tau_{\pm}$ as
the
integration variables $\varphi\,(\tau)$ in (\ref{eqn:6.3}) and the
functional
composition law $\stackrel{\leftrightarrow} {\delta{\mbox{\boldmath
$F$}}}\!{\mbox{\boldmath $G$}}$ implies a symmetric action of time
derivatives
on both arguments of ${\mbox{\boldmath $G$}}\,(\tau,\tau^{\prime}
\!)$ similar
to (\ref{eqn:6.4}):
         \begin{eqnarray}
	 &&{\rm Tr}\,\stackrel{\leftrightarrow}
	 {\delta{\mbox{\boldmath $F$}}}\!{\mbox{\boldmath
$G$}}=\int_{\tau_{-}}^{\tau_{+}}d\tau\,
	 {\rm tr}\left[\,\stackrel{\leftrightarrow}
	 {\delta{\mbox{\boldmath $F$}}}\!{\mbox{\boldmath
$G$}}\,(\tau,\tau^{\prime}\!)\,\right]_
	 {\tau^{\prime}=\tau} \nonumber \\
	 \nonumber \\
	 &&\;\;\;\;\;\;\;\;\;\;\;\;\;\;\;\;
	 \equiv\int_{\tau_{-}}^{\tau_{+}}d\tau\,
	 {\rm tr}\left[\,(\delta a\,\frac{d^{2}}{d\tau\,
	 d\tau^{\prime}}+\delta b\,\frac{d}{d\tau^{\prime}}+
	 \delta b^{T}\,\frac{d}{d\tau}+\delta c)\,
	 {\mbox{\boldmath $G$}}\,(\tau,\tau^{\prime}\!)\,
	 \right]_{\tau^{\prime}=\tau}.
\label{eqn:6.8}
	 \end{eqnarray}
Here ${\rm tr}$ denotes the matrix trace operation with respect to
condensed
indices of $\delta a=\delta a_{ik},\,\delta b=\delta b_{ik},\,\delta
b^{T}=\delta b_{ki},\,\delta c=\delta c_{ik}$ and ${\mbox{\boldmath
$G$}}\,(\tau,\tau^{\prime}\!)={\mbox{\boldmath $G$}}^{ki}\,
(\tau,\tau^{\prime}\!)$. As discussed above, in field theory this
trace has
also a functional nature because it involves either a spatial
integration or
infinite summation over quantum numbers of spatial harmonics, but its
functional dimensionality is lower than in ${\rm Tr}$ for it does not
involve
integration over time.

The purpose of this paper will be to integrate the variational
equation
(\ref{eqn:6.8}) and, thus, obtain the closed algorithm for the
functional
determinant ${\rm Det}\,{\mbox{\boldmath $F$}}$ in terms of
quantities of lower
functional dimensionality (involving the matrix trace (tr) and
determinant
(det) operations on condensed indices $i,k,...$ of the above type).
We shall do
it for three different types of boundary conditions  in the gaussian
path
integral (\ref{eqn:6.3}). The first type corresponds to the
calculation of the
kernel (\ref{eqn:2.2}) of transition between two {\it regular}
hypersurfaces
$\Sigma_{\pm}$ of finite size at $\tau_{\pm}$ (see Fig.1) with fixed
"end
point" configurations $\phi_{\pm}$, which imply the Dirichlet
boundary
conditions on quantum inegration variables $\varphi(\tau_{\pm}\!)=0$.
The
second type incorporates the calculation of the wavefunction of
Hartle and
Hawking with fixed field at $\tau_{+}$, $\varphi(\tau_{+})=0$, and
the
so-called no-boundary condition at $\tau_{-}$. In this case the
integration
goes over all $\varphi(\tau_{-})$ satisfying the necessary regularity
properties providing that  $\tau_{-}$ is a regular inner point of the
Euclidean
spacetime which has as the only boundary the surface $\Sigma_{+}$ at
$\tau_{+}$. Finally, we consider the third type incorporating the
"no-boundary"
conditions at the both ends of time segment $\tau=\tau_{\pm}$ and
corresponding
to the one-loop preexponential factor of the theory on the closed
Euclidean
spacetime of spherical topology. The first type represents the
path-integral
derivation of the well-known Pauli - Van Vleck - Morette formula
\cite{Pauli}
for the kernel of the heat equation, and we shall begin our
considerations in
the next section with this case.

The final remark, which is in order here, concerns the ultraviolet
infinities
in the one-loop preexponential factors inalienable in any local field
theory
and necessarily arising in the transition from quantum mechanical to
field
models with infinite number of degrees of freedom. In this paper we
shall not
consider this problem implying that any kind of ultraviolet
regularization can
be performed in both sides of the reduction algorithms which we are
going to
obtain. The justification of this strategy was partly explained in
the previous
paper of this series \cite{tunnelI} (see Sect.8 of that paper) and
based on the
fact that these reduction algorithms, beeing intrinsically
non-covariant, serve
as a bridge between the non-covariant manifestly unitary Lorentzian
theory and
its Euclidean version accumulating in a manifestly covariant form all
divergent
quantum corrections. It is the latter theory which must be
covariantly
regulated to give physically reasonable results, in contrast to the
ultraviolet
regularization of non-covariant quantities \cite{zeta,BKK,BKKM} which
may lead
to the discrepancies with their manifestly covariant counterparts
\cite{Griffin}
\footnote
{See the disagreement between the covariant calculation of
gravitational
$\zeta$-functions in \cite{covzeta} and their non-covariant
calculation in
terms of physical variables \cite{Griffin,BKK,BKKM}, which implies
that the
unification of the covariant Euclidean quantum gravity and its
Lorentzian
unitary counterpart requires a deeper consideration accounting for
subtleties
of the ultraviolet regularization in covariant and unitary gauges.}.

\section{The Pauli - Van Vleck - Morette formula for the one-loop
preexponential factor}
\hspace{\parindent}
Let us consider the operator (\ref{eqn:3.22}) on a finite segment of
the
Euclidean time $\tau_{-}\leq\tau\leq\tau_{+}$ such that all its
coefficients
$a,\,b$ and $c$ are regular functions on this segment. This operator
is
formally symmetric under the operation of integration by parts (and
matrix
transposition with respect to indices $i$) and, therefore, it is
selfadjoint
under boundary conditions in the space of fields, which provide the
vanishing
of the total-derivative terms. The boundary conditions on the
integration
variables $\varphi(\tau_{\pm}\!)=0$ fall into this category and
correspond to
the calculation of the one-loop contribution for the Euclidean
transition
kernel from $\tau_{-}$ to $\tau_{+}$.

For the integration of eq.(\ref{eqn:6.7}) let us introduce two sets
of basis
functions ${\mbox{\boldmath $u_{-}$}}$ and ${\mbox{\boldmath
$u_{+}$}}$ of the
operator ${\mbox{\boldmath $F$}}$
          \begin{equation}
	  {\mbox{\boldmath $F$}}{\mbox{\boldmath $u_{\pm}$}}=0,\,\,
	  {\mbox{\boldmath $u_{\pm}$}}={\mbox{\boldmath
$u_{\pm}$}}^{i}_{A}(\tau)
  \label{eqn:6.9}
	  \end{equation}
satisfying the Dirichlet boundary conditions respectively at
$\tau_{-}$ and
$\tau_{+}$:
          \begin{equation}
	  {\mbox{\boldmath $u_{-}$}}(\tau_{-}\!)=0,\,\,
	  {\mbox{\boldmath $u_{+}$}}(\tau_{+}\!)=0.
\label{eqn:6.10}
	  \end{equation}
In the DeWitt notations, which will be used throughout the paper, we
regard
these basis functions, enumerated by the condensed index $A$ of
arbitrary
nature, as forming the square matrices with the {\it first}
(contravariant)
index $i$ and the {\it second} (covariant) index $A$. According to
the
discussion of the previous paper \cite{tunnelI}, containig examples
of such
functional matrix structure, these matrices can be regarded
invertible at
general $\tau$ except certain particular values of time including
$\tau_{-}$
for ${\mbox{\boldmath $u_{-}$}}$ and $\tau_{+}$ for ${\mbox{\boldmath
$u_{+}$}}$. As is shown in the Appendix, if the operator
${\mbox{\boldmath
$F$}}$ does not have zero egenvalues at the segment
$\tau_{-}\leq\tau\leq\tau_{+}$, these matrices are granted to be
invertible at
the ends of this segment opposite to the points of their boundary
conditions
(i.e. ${\mbox{\boldmath $u_{\pm}$}}\,(\tau_{\mp})$) and can be used
to form
other two {\it invertible} matrices related by a simple transposition
law
          \begin{eqnarray}
	  &&{\mbox{\boldmath $\Delta$}}_{+-}={\mbox{\boldmath
$u^{T}_{+}$}}\,
	  ({\mbox{\boldmath $W$}}{\mbox{\boldmath $u$}}_{-}\!)-
	  ({\mbox{\boldmath $W$}}{\mbox{\boldmath
$u$}}_{+}\!)^{T}{\mbox{\boldmath
$u$}}_{-},\,\,
	  {\mbox{\boldmath $\Delta$}}_{+-}
	  \equiv({\mbox{\boldmath $\Delta$}}_{+-}\!)_{AB},
\label{eqn:6.11}
	  \\
	  &&{\mbox{\boldmath $\Delta$}}_{-+}=
	  {\mbox{\boldmath $u^{T}_{-}$}}\,({\mbox{\boldmath
$W$}}{\mbox{\boldmath
$u$}}_{+}\!)-
	  ({\mbox{\boldmath $W$}}{\mbox{\boldmath
$u$}}_{-}\!)^{T}{\mbox{\boldmath
$u$}}_{+},\,\,
	  {\mbox{\boldmath $\Delta$}}_{-+}
	  \equiv({\mbox{\boldmath $\Delta$}}_{-+}\!)_{AB},
\label{eqn:6.12}
	  \\
	  &&{\mbox{\boldmath $\Delta$}}_{+-}^{T}
	  =-{\mbox{\boldmath $\Delta$}}_{-+}.
\label{eqn:6.13}
	  \end{eqnarray}
In view of the Wronskian relation (\ref{eqn:6.6}) for the operator
${\mbox{\boldmath $F$}}$ these matrices are constants of motion
 	  \begin{equation}
	  \frac{d}{d\tau}{\mbox{\boldmath $\Delta$}}_{+-}=0,\,\,
	  \frac{d}{d\tau}{\mbox{\boldmath $\Delta$}}_{-+}=0
\label{eqn:6.14}
	  \end{equation}
and enter the following important relations  for equal-time bilinear
combinations of basis functions ${\mbox{\boldmath
$u_{\pm}$}}={\mbox{\boldmath
$u_{\pm}$}}(\tau)$ (see Appendix)
          \begin{eqnarray}
	  &&{\mbox{\boldmath $u$}}_{+}(\tau)\,
	  ({\mbox{\boldmath $\Delta$}}_{-+}\!)^{-1}{\mbox{\boldmath
$u$}}_{-}^{T}(\tau)
	  +{\mbox{\boldmath $u$}}_{-}(\tau)\,({\mbox{\boldmath
$\Delta$}}_{+-}\!)^{-1}
	  {\mbox{\boldmath $u$}}_{+}^{T}(\tau)=0,
\label{eqn:6.15}
	  \\
	  &&a\left[\,\dot{{\mbox{\boldmath $u$}}}_{+}(\tau)\,
	  ({\mbox{\boldmath $\Delta$}}_{-+}\!)^{-1}
	  {\mbox{\boldmath $u$}}_{-}^{T}(\tau)
	  +\dot{{\mbox{\boldmath $u$}}}_{-}(\tau)\,
	  ({\mbox{\boldmath $\Delta$}}_{+-}\!)^{-1}
	  {\mbox{\boldmath
$u$}}_{+}^{T}(\tau)\,\right]={\mbox{\boldmath $I$}}.
 \label{eqn:6.16}
	  \end{eqnarray}
Here ${\mbox{\boldmath $I$}}=\delta^{i}_{k}$ denotes the unity matrix
in the
space of indices $i$. Just to clarify the use of indices in these
equations and
in what follows, note that in the transposed  matrix
${\mbox{\boldmath
$u$}}^{T}_{\pm}= {\mbox{\boldmath $u_{\pm}$}}_{A}^{i}$ the covariant
index $A$
is considered to be the first one (in contrast to ${\mbox{\boldmath
$u$}}_{\pm}$), so that the matrix composition law with
$({\mbox{\boldmath
$\Delta$}}_{+-}\!)^{-1}= [\,({\mbox{\boldmath
$\Delta$}}_{+-}\!)^{-1}\,]^{AB}$
gives rise in eqs.(\ref{eqn:6.15}) to the matrices with two field
indices $i$
and $k$.

The basis functions and their Wronskian matrices (\ref{eqn:6.11}) -
(\ref{eqn:6.12}) can be used to construct the following expression
for the
Green's function of the operator ${\mbox{\boldmath $F$}}$ subject to
Dirichlet
boundary conditions
           \begin{eqnarray}
	   &\!\!\!\!\!\!{\mbox{\boldmath
$G$}}\,(\tau,\tau^{\prime}\!)=
	   -&\!\!\!\!\!\!\theta\,(\tau\!-\!\tau^{\prime}\!)\,
	   {\mbox{\boldmath $u$}}_{\!+}(\tau)\,
	   ({\mbox{\boldmath $\Delta$}}_{-+}\!)^{-1}{\mbox{\boldmath
$u$}}_{-}^{T}
	   (\tau^{\prime}\!)+\theta\,(\tau^{\prime}\!-\!\tau)\,
	   {\mbox{\boldmath $u$}}_{\!-}(\tau)\,({\mbox{\boldmath
$\Delta$}}_{+-}\!)^{-1}
	   {\mbox{\boldmath $u$}}_{+}^{T}(\tau^{\prime}\!),
\label{eqn:6.17a}
	   \\
	   &\!\!\!\!{\mbox{\boldmath $G$}}\,(\tau_{\pm},
\label{eqn:6.17}
	   \tau^{\prime}\!)=0,\!&
	   \end{eqnarray}
where $\theta(x)$ represents the step function: $\theta(x)=1$ for
$x\geq 0$ and
$\theta(x)=0$ for $x<0$. This expression, the derivation of which is
presented
in Appendix, is nothing but the analogue of positive-negative
frequency
decomposition for the Feynman propagator in asymptotically-flat
spacetime
\cite{DW:Dynamical}: ${\mbox{\boldmath $u$}}_{\pm}$ play the role of
positive
and negative frequency basis functions of the Lorentzian Klein-Gordon
equation
and the boundary conditions (\ref{eqn:6.10}) serve as a Euclidean
counterpart
to the IN-OUT Feynman boundary conditions propagating positive
frequencies to
the future and negative ones to the past
\footnote
{Note that the Wick rotation of Lorentzian positive and negative
frequency
basis functions makes them vanish respectively at the "future" and
"past"
infinity of the Euclidean time which corresponds to the
$\tau_{\pm}\rightarrow
\pm\infty$ limit in the boundary conditions (\ref{eqn:6.10}), so that
they can
be regarded as a generalization of the Euclidean asymptotically-flat
boundary
conditions to the case of a finite time interval.
}.

Now we can use eq.(\ref{eqn:6.17a}) to calculate the variational
functional
trace (\ref{eqn:6.8}). The action of derivatives on ${\mbox{\boldmath
$G$}}\,(\tau, \tau^{\prime}\!)$ in this trace gives rise to
additional terms
containing the coincidence limit of the delta function
$\delta\,(\tau-\tau^{\prime}\!)
=(d/dt)\theta\,(\tau\!-\!\tau^{\prime}\!)$ and
its first order derivative $(d/dt)\delta\,(\tau-\tau^{\prime}\!)$. It
is
remarkable that the total coefficient of
$(d/dt)\delta\,(\tau-\tau^{\prime}\!)$
vanishes in virtue of the eq.(\ref{eqn:6.15}), while the rest of the
terms give
the expression
         \begin{eqnarray}
	 {\rm Tr}\,\stackrel{\leftrightarrow}
	 {\delta{\mbox{\boldmath $F$}}}\!{\mbox{\boldmath
$G$}}&\!\!\!=&\!\!\!
	 \int_{\tau_{-}}^{\tau_{+}}d\tau\,
	 \delta\,(0)\;{\rm tr}\,\delta a\,
	 \left[\,\dot{{\mbox{\boldmath $u$}}}_{+}
	 ({\mbox{\boldmath $\Delta$}}_{-+}\!)^{-1}
	 {\mbox{\boldmath $u$}}_{-}^{T}+\dot{{\mbox{\boldmath
$u$}}}_{-}
	 ({\mbox{\boldmath $\Delta$}}_{+-}\!)^{-1}
	 {\mbox{\boldmath $u$}}_{+}^{T}\,\right]
\label{eqn:6.18}
	 \\
	 &\!\!\!-&\!\!\!\int_{\tau_{-}}^{\tau_{+}}d\tau\,
	 {\rm tr}\;\left\{\,\theta\,(\tau\!-\!\tau^{\prime})\,
	 ({\mbox{\boldmath $\Delta$}}_{-+}\!)^{-1}{\mbox{\boldmath
$u$}}_{-}^{T}
	 \stackrel{\leftrightarrow}{\delta{\mbox{\boldmath $F$}}}
	 {\mbox{\boldmath $u$}}_{+}-
	 \theta\,(\tau^{\prime}\!-\!\tau)\,
	 ({\mbox{\boldmath $\Delta$}}_{+-}\!)^{-1}{\mbox{\boldmath
$u$}}_{+}^{T}
	 \stackrel{\leftrightarrow}{\delta{\mbox{\boldmath $F$}}}
	 {\mbox{\boldmath $u$}}_{-}\right\}_{\tau^{\prime}=\tau}.
\nonumber
	 \end{eqnarray}
Notice that in view of (\ref{eqn:6.16}) the coefficient of $\delta a$
equals
$a^{-1}$, so that the first integral on the right-hand side of this
equation
reduces to the variation of the logarithm of local measure
(\ref{eqn:3.6}). On
the other hand, the symmetry of the operator
$\stackrel{\leftrightarrow}
{\delta{\mbox{\boldmath $F$}}}$, defined by the eq.(\ref{eqn:6.4})
with the
coefficients $a,\,b$ and $c$ replaced by their variations, together
with the
transposition property (\ref{eqn:6.13}) shows that the expression in
the second
integral reduces to the form free from discontinuous
$\theta$-functions and,
thus, has a well-defined coincidence limit. Therefore, combining
(\ref{eqn:6.7}), (\ref{eqn:6.18}) and (\ref{eqn:3.6}) one arrives at
the
following expression for the variation of functional determinants in
the
one-loop preexponential factor of (\ref{eqn:3.18})
         \begin{equation}
	 \delta\;{\rm ln}\,\frac{{\rm Det}\,{\mbox{\boldmath $F$}}}
	 {{\rm Det}\,{\mbox{\boldmath $a$}}}=
	 -\int_{\tau_{-}}^{\tau_{+}}d\tau\,
	 {\rm tr}\;({\mbox{\boldmath
$\Delta$}}_{-+}\!)^{-1}{\mbox{\boldmath
$u$}}_{-}^{T}
	 \stackrel{\leftrightarrow}{\delta{\mbox{\boldmath $F$}}}
	 {\mbox{\boldmath $u$}}_{+}.
\label{eqn:6.19}
	 \end{equation}
Here we reserve the same notation tr for the trace operation in the
space of
indices $A$ enumerating the basis functions of ${\mbox{\boldmath
$F$}}$, so
that the integrand of (\ref{eqn:6.19}) should read
$[\,({\mbox{\boldmath
$\Delta$}}_{-+}\!)^{-1}\,]^{AB} {\mbox{\boldmath
$u$}}_{-B}^{T}\stackrel{\leftrightarrow}{\delta{\mbox{\boldmath
$F$}}}{\mbox{\boldmath $u$}}_{+A}$.

The above equation reveals the role of the local measure
(\ref{eqn:3.6}) in the
 one-loop approximation of quantum theory: it exactly cancels a
strongest power
divergent part in the functional determinant of the operator of
linearized
quantum disturbancies. This result was obtained for general dynamical
systems
by DeWitt \cite{DW:Dynamical,DW:LesH} and studied for covariant field
theories
by Fradkin and Vilkovisky \cite{FV:Bern} with a special emphasis on
peculiarities of the ultraviolet regularization for functional traces
\footnote
{In field theories the cancellation of the leading volume
divergencies by the
contribution of local measure is intertwined with the ultraviolet
regularization, but in the main it has the same origin as above:
Green's
function  has a nature of a singular generalized function which
generates the
power divergencies in the functional trace when it is acted upon by
the second
order differential operator.}.
The right-hand side of (\ref{eqn:6.19}) does not contain the
$\delta(0)$-type
divergence (and simply finite for quantum mechanical systems with
finite-dimensional configuration space). As we shall show now, it can
be
further simplified to the expression not involving the integration
over time
and, thus, effectively reducing the dimensionality of the functional
determinant and trace.

Let us use in the right-hand side of (\ref{eqn:6.19}) the equation
obtained
from varying the operators ${\mbox{\boldmath $F$}}$ and
${\mbox{\boldmath
$W$}}$ in the identity (\ref{eqn:6.5}) with a subsequent substitution
$\varphi^{T}_{1}= {\mbox{\boldmath $u$}}_{-}^{T}$ and
$\varphi_{2}={\mbox{\boldmath $u$}}_{+}$:
	   \begin{equation}
	   {\mbox{\boldmath $u$}}_{-}^{T}\!\stackrel
	   {\leftrightarrow}{\delta{\mbox{\boldmath
$F$}}}\!{\mbox{\boldmath
$u$}}_{+}=
	   {\mbox{\boldmath $u$}}_{-}^{T}\,(\delta{\mbox{\boldmath
$F$}}{\mbox{\boldmath $u$}}_{+}\!)+
	   \frac{d}{d\tau}\left[\,{\mbox{\boldmath $u$}}_{-}^{T}\,
	   (\delta{\mbox{\boldmath $W$}}\!{\mbox{\boldmath
$u$}}_{+}\!)\,\right].
  \label{eqn:6.20}
	   \end{equation}
We also have the varied version of eq.(\ref{eqn:6.9}) which forms the
following
boundary-value problem for the variations of basis functions
	   \begin{eqnarray}
	   &&{\mbox{\boldmath $F$}}\delta{\mbox{\boldmath
$u$}}_{\pm}=
	   -\delta{\mbox{\boldmath $F$}}{\mbox{\boldmath
$u$}}_{\pm},\\
 \label{eqn:6.21}
	   &&\delta{\mbox{\boldmath $u$}}_{+}(\tau_{+}\!)=0,\,\,
	   \delta{\mbox{\boldmath $u$}}_{-}(\tau_{-}\!)=0.
\label{eqn:6.22}
	   \end{eqnarray}
This allows to rewrite the first term on the right hand side of
(\ref{eqn:6.20}) as ${\mbox{\boldmath
$u$}}_{-}^{T}\,(\delta{\mbox{\boldmath
$F$}}{\mbox{\boldmath $u$}}_{+}\!)= -{\mbox{\boldmath
$u$}}_{-}^{T}\,({\mbox{\boldmath $F$}}\delta{\mbox{\boldmath
$u$}}_{+}\!)$ and
then use the Wronskian relation (\ref{eqn:6.6}) with
$\varphi^{T}_{1}=
{\mbox{\boldmath $u$}}_{-}^{T}$ and
$\varphi_{2}=\delta{\mbox{\boldmath
$u$}}_{+}$. Therefore, the integrand in the right hand side of
(\ref{eqn:6.19})
reduces to a total derivative and yields the contribution of surface
terms at
$\tau_{\pm}$
	 \begin{equation}
	 \delta\;{\rm ln}\,\frac{{\rm Det}\,{\mbox{\boldmath $F$}}}
	 {{\rm Det}\,{\mbox{\boldmath $a$}}}=
	 -{\rm tr}\;({\mbox{\boldmath
$\Delta$}}_{-+}\!)^{-1}\left[\left.{\mbox{\boldmath $u$}}_{-}^{T}\,
	 \delta\,({\mbox{\boldmath $W$}}\!{\mbox{\boldmath
$u$}}_{+}\!)\,\right|_{\tau_{+}}\!\!+
	 \left.({\mbox{\boldmath $W$}}\!{\mbox{\boldmath
$u$}}_{-}\!)^{T}\delta
	 {\mbox{\boldmath $u$}}_{+}\,\right|_{\tau_{-}}\right].
\label{eqn:6.23}
	 \end{equation}
Here the other two surface terms vanish in virtue of the boundary
conditions
for basis functions and the expression $\delta\,({\mbox{\boldmath
$W$}}\!{\mbox{\boldmath $u$}}_{+}\!)=
(\delta{\mbox{\boldmath $W$}}\!){\mbox{\boldmath
$u$}}_{+}+{\mbox{\boldmath
$W$}}(\delta{\mbox{\boldmath $u$}}_{+}\!)$ involves the variations of
both the
Wronskian operator and the basis function with respect to arbitrary
variations
of coefficients in the operator ${\mbox{\boldmath $F$}}$.

On the other hand, in virtue of the same boundary conditions the
conserved
matrix (\ref{eqn:6.12}), ${\mbox{\boldmath $\Delta$}}_{-+}$, when
evaluated at
$\tau_{+}$ and $\tau_{-}$, has two different representations
	\begin{eqnarray}
	&&{\mbox{\boldmath $\Delta$}}_{-+}=\,{\mbox{\boldmath
$u$}}_{-}^{T}\,
	({\mbox{\boldmath $W$}}\!{\mbox{\boldmath
$u$}}_{+}\!)\;|_{\tau_{+}},
\label{eqn:6.24}\\
	&&{\mbox{\boldmath $\Delta$}}_{-+}=-({\mbox{\boldmath
$W$}}\!{\mbox{\boldmath
$u$}}_{-}\!)^{T}
	{\mbox{\boldmath $u$}}_{+}\,|_{\tau_{-}}.
\label{eqn:6.25}
	\end{eqnarray}
They allow us to convert the right hand side of (\ref{eqn:6.23}) into
the total
variation
	 \begin{eqnarray}
	 \delta\;{\rm ln}\,\frac{{\rm Det}\,{\mbox{\boldmath $F$}}}
	 {{\rm Det}\,{\mbox{\boldmath $a$}}}=
	 \!-{\rm tr}\left[\,\left.({\mbox{\boldmath
$W$}}\!{\mbox{\boldmath
$u$}}_{+}\!)^{-1}
	 \delta\,({\mbox{\boldmath $W$}}\!{\mbox{\boldmath
$u$}}_{+}\!)\;\right|_{\tau_{+}}\!\!\!-
	 \left.{\mbox{\boldmath $u$}}_{+}^{-1}\delta
	 {\mbox{\boldmath $u$}}_{+}\right|_{\tau_{-}}\right]\!
	 =\!-\delta\,{\rm ln\,det}\,I_{+-}
\label{eqn:6.26}
	 \end{eqnarray}
of the logarithm of the determinant of the following matrix
$I_{+-}=(I_{+-})_{ik}$
	\begin{equation}
	I_{+-}=({\mbox{\boldmath $W$}}\!{\mbox{\boldmath
$u$}}_{+}\!)\,(\tau_{+})\,
	{\mbox{\boldmath $u$}}_{+}^{-1}(\tau_{-})=
	{\mbox{\boldmath $u$}}^{T-1}_{-}(\tau_{+})\,
	{\mbox{\boldmath $\Delta$}}_{-+}\,{\mbox{\boldmath
$u$}}_{+}^{-1}(\tau_{-}).
\label{eqn:6.27}
	\end{equation}
Now the functional integration of the equation (\ref{eqn:6.26})
presents no
difficulty and gives the following answer for the one-loop
preexponential
factor
	\begin{eqnarray}
	\left(\,\frac{{\rm Det}\,{\mbox{\boldmath $F$}}}{{\rm Det}\,
	{\mbox{\boldmath $a$}}}\,\right)^{-1/2}=
	{\rm Const}\,\left(\,{\rm det}\,
	I_{+-}\right)^{1/2},              \label{eqn:6.28}
	\end{eqnarray}
where Const represents a completely field independent normalization
constant.

This equation reduces the functional dimensionality of the
determinant from Det
to det and represents a well-known Pauli-Van Vleck-Morette formula
for the
one-loop preexponential factor in the heat kernel or, in its
Lorentzian
version, the kernel of the unitary Schrodinger evolution
\cite{Pauli}. Under a
certain Hermitian operator realization of the Hamiltonian $\hat H$ in
the heat
equation (\ref{eqn:6.29}), this formula for
$K(\tau_{+},\phi_{+}|\tau_{-},
\phi_{-}\!)$ reads
	\begin{eqnarray}
	K(\tau_{+},\phi_{+}|\,\tau_{-},\phi_{-}\!)=
	\left[\,{\rm det}\,\left(-\frac{1}{2\pi\hbar}\,\right)
	\frac{\partial^{2}I}
	{\partial\phi_{+}\partial\phi_{-}}\,
	\right]^{1/2}{\rm e}^{\!\!\phantom{0}^{\textstyle
        -\frac{1}{\hbar}I}}
	\,\left[\,1+O\,(\,\hbar\,)\,\right],     \label{eqn:6.31}
        \end{eqnarray}
where
	\begin{equation}
	I\equiv I\,(\tau_{+},\phi_{+}|\,\tau_{-},\phi_{-}\!)=
	I\,[\,\phi\,]\,|_
	{\,\phi\,(\tau)=\phi\,(\tau,\phi_{\pm})}
\label{eqn:6.31a}
	\end{equation}
is the Euclidean Hamilton-Jacobi function of the theory -- the
classical action
calculated at the extremal of the equations of motion
(\ref{eqn:3.18a}) and
parametrized in terms of its end points on the segment of the
Euclidean time
$[\,\tau_{-},\tau_{+}]$
\footnote
{The formula (\ref{eqn:6.31}), which is much better known in the
Lorentzian
context with $I=-iS$, can be obtained by solving in the semiclassical
approximation the heat equation (\ref{eqn:6.29}). Such a general
derivation was
recently presented in \cite{BarvU}, where the precise operator
realization of
$\hat H$ corresponding to (\ref{eqn:6.31}) (and usually disregarded
in physics
literature) was formulated.
}.
The comparison of (\ref{eqn:6.31}) with (\ref{eqn:3.18}) shows that
the matrix
$\partial^{2}I/\partial\phi_{+}\partial\phi_{-}$ should coincide with
the
matrix (\ref{eqn:6.27}) in the just obtained algorithm
(\ref{eqn:6.28}) for the
one-loop preexponential factor. To prove this property, note that
this matrix
can be written as a variation with respect to $\phi_{-}$ of the
Euclidean
momentum of this extremal at $\tau_{+}$, $\partial
I/\partial\phi_{+}=\partial{\cal L}_{E}/\partial\dot\phi(\tau_{+})$.
Therefore,
in virtue of the variational relation (\ref{eqn:4.8}) the above
matrix takes
the form
	\begin{equation}
	\frac{\partial^{2}I}
	{\partial\phi_{+}\partial\phi_{-}}=
	\left.{\mbox{\boldmath $W$}}(d/d\tau\!)\;
	\frac{\partial\phi(\tau,\phi_{\pm}\!)}
	{\partial\phi_{-}}\;\right|_{\,\tau=\tau_{+}}.
\label{eqn:6.33}
	\end{equation}
On the other hand, the Jacobi matrix
$\partial\phi(\tau|\phi_{\pm}\!)/\partial
\phi_{-}$ satisfies the linear boundary value problem, derivable by
varying the
classical equations \ref{eqn:3.18a}) with respect to the boundary
data
$\phi_{-}$,
	\begin{eqnarray}
	&&{\mbox{\boldmath
$F$}}(d/d\tau\!)\;\partial\phi(\tau,\phi_{\pm}\!)
	/\partial\phi_{-}=0,\nonumber \\
	&&\partial\phi(\tau_{+},\phi_{\pm}\!)
	/\partial\phi_{-}=0,\,
	\partial\phi(\tau_{-},\phi_{\pm}\!)
	/\partial\phi_{-}={\mbox{\boldmath $I$}}
\label{eqn:6.34}
	\end{eqnarray}
and, therefore, has the following decomposition in terms of the basis
functions
${\mbox{\boldmath $u$}}_{+}(\tau)$ of the operator ${\mbox{\boldmath
$F$}}$
	\begin{eqnarray}
	\partial\phi(\tau,\phi_{\pm}\!)
	/\partial\phi_{-}={\mbox{\boldmath $u$}}_{+}(\tau)\,
	{\mbox{\boldmath $u$}}_{+}^{-1}(\tau_{-})
\label{eqn:6.35}
	\end{eqnarray}
whence we get the equality of matrices (\ref{eqn:6.27}) and
(\ref{eqn:6.33})
and the needed equivalence of the reduction algorithm
(\ref{eqn:6.28}) to the
Pauli-Van Vleck-Morette formula (\ref{eqn:6.31}).

\section{The functional determinant on spacetime of the no-boundary
type}
\hspace{\parindent}
The derivations of the previous section can be now applied to the
case of
primary interest in this paper -- the calculation of the functional
determinant
subject to the no-boundary regularity conditions. All the functions
above were
assumed to be regular on the time segment
$\tau_{-}\leq\tau\leq\tau_{+}$
corresponding in field theory to the calculations on Euclidean
spacetime
sandwiched between two regular spatial slices $\Sigma_{\pm}$. For
spatially
closed cosmology with $\Sigma=S^{3}$ such a spacetime has a topology
of a tube
$[\,\tau_{-},\tau_{+}] \times S^{3}$, and, as it was discussed in
\cite{tunnelI}, a no-boundary spacetime can be obtained from this
tube by
shrinking the surface $\Sigma_{-}=S^{3}_{-}$ to a point (see Fig.2),
so that
the variable $\tau-\tau_{-}=\tau$ in the resulting four-dimensional
ball plays
the role of a radial coordinate (for simplicity in this section we
shall assume
that $\tau_{-}=0$) and the metric in the vicinity of $\tau_{-}$ --
the regular
internal point of the Euclidean manifold -- takes the form
	\begin{equation}
	ds^{2}=d\tau^{2}+
	\tau^{2}c_{ab}\,dx^{a}dx^{b}+O(\tau^{3}),\;\;
	\tau\rightarrow 0,
\label{eqn:2.20}
        \end{equation}
with $c_{ab}$ - the round metric of a three-dimensional sphere of
unit radius,
parametrized with some quasi-angular coordinates $x^{a}$. The
boundary
conditions for the one-loop functional determinant coincide on the
boundary of
this spacetime, $\tau=\tau_{+}$, with those of the previous section,
but in the
center of this ball they drastically differ, because instead of a
Dirichlet
type they correspond to the conditions of regularity. The latter are
rather
nontrivial because the center of the spacetime ball is a singular
point of the
radial part of the differential operator ${\mbox{\boldmath
$F$}}(d/d\tau\!)$,
which results in a special behaviour of basis functions
${\mbox{\boldmath
$u$}}_{\pm}(\tau)$ in the vicinity of $\tau_{-}$.

Indeed, the linearized physical modes of all possible spins,
$s=0,1/2,1,3/2,2,...$, are just the transverse-traceless components
$\phi^{i}=(\varphi,\,\phi({\bf x}),\,\psi({\bf x}),\,
A^{T}_{a}({\bf x}),\,\psi^{T}_{a}({\bf x}),\,h^{TT}_{ab}({\bf
x}),...)$ of the
three-dimensional tensor fields in the $\tau$-foliation of the
Euclidean
spacetime \cite{tunnelI}. For them, the coefficient $a=a_{ik}$ of the
second-order derivatives in ${\mbox{\boldmath $F$}}(d/d\tau\!)$ can
be
collectively written as
	\begin{eqnarray}
	a_{ik}=(^{4}g)^{1/2}g^{\tau\tau}g^{a_{1}a_{2}}...
	g^{a_{s}a_{2s}}\,\delta\,({\bf x}_{i}-{\bf x}_{k}),
\label{eqn:6.2.1}
	\end{eqnarray}
$i=(a_{1},...a_{s},{\bf x}_{i}),\,k=(a_{2},...a_{2s},{\bf x}_{k}),$
and in the
regular metric (\ref{eqn:2.20}) has the following asymptotic
behaviour
	\begin{equation}
	a=a_{0}\,\tau^{k}+O\,(\,\tau^{k+1}\,),\,\,
	k=3-2s,\,\,\tau\rightarrow\tau_{-}=0,
\label{eqn:6.2.2}
	\end{equation}
where $a_{0}$ is defined by eq.(\ref{eqn:6.2.1}) with respect to the
round
metric $c_{ab}$ on a 3-sphere of the unit radius and the unit lapse
$g^{\tau\tau}=N^{-2}=1$
\footnote
{Even though the expression (\ref{eqn:6.2.1}) is formally valid only
for
integer-spin fields, this behaviour with the parameter $k=3-2s$ also
holds for
half-integer spins, because every  gamma matrix substituting the
corresponding
metric coefficient in (\ref{eqn:6.2.1}) contributes one power of
$\tau^{-1}$.}
{}.
Therefore the equations (\ref{eqn:6.9}) for basis functions have the
form
	\begin{equation}
	\left(\frac{d^2}{d\tau^2}+f\frac{d}{d\tau}+
	g\right){\mbox{\boldmath $u$}}_{\pm}(\tau)=0,
\label{eqn:6.2.3}
	\end{equation}
with the coefficients $f$ and $g$ possessing the following asymptotic
behaviour
	\begin{equation}
	f=\frac{k}{\tau}\,{\mbox{\boldmath $I$}}+
	O\,(\,\tau^{0}\,),\,\,\,
	g=\frac{g_{\,0}}{\tau^2}+
	O\,(\,\tau^{-1}\,).                   \label{eqn:6.2.4}
	\end{equation}
Here the leading singularity in the potential term $g$ originates
from the
spatial Laplacian $g^{ab}\nabla_{a}\nabla_{b}$ entering the operator
${\mbox{\boldmath $F$}}$, which scales in the metric (\ref{eqn:2.20})
as
$1/\tau^{2}$, and the leading term of $f$ is always a multiple of the
unity
matrix {\mbox{\boldmath $I$}} with the same parameter $k=3-2s$ as in
(\ref{eqn:6.2.2}). In the representation of spatial harmonics, the
eigenfunctions of a spatial Laplacian, the (functional) matrix
$g_{\,0}$ can be
also diagonalized, $g_{\,0}={\rm diag}\{-\omega^2_{i}\}$ , so that,
without the
loss of generality, the both singularities in (\ref{eqn:6.2.3}) can
be
characterised by simple numbers $k$ and $\omega^2=\omega^2_{i}$ for
every
component of ${\mbox{\boldmath $u$}}_{\pm}={\mbox{\boldmath
$u$}}_{\pm}^{i}$.

As it follows from the theory of differential equations with singular
points
\cite{Olver}, in this case the asymptotic behaviour of the solutions
${\mbox{\boldmath $u$}}_{\pm}(\tau)$ near $\tau_{-}=0$ has the form
	\begin{eqnarray}
	{\mbox{\boldmath $u$}}_{-}(\tau)={\mbox{\boldmath
$U$}}_{\!-}\,
	\tau^{\mu_{-}}
	+O\,(\,\tau^{1+\mu_{-}}),      \label{eqn:6.2.5}\\
	{\mbox{\boldmath $u$}}_{+}(\tau)={\mbox{\boldmath
$V$}}_{\!+}\,
	\tau^{\mu_{+}}
	+O\,(\;\tau^{1+\mu_{+}}),        \label{eqn:6.2.6}
	\end{eqnarray}
where $\mu_{\pm}$ are the roots of the following quadratic equation
involving
only the coefficients of leading singularities
	\begin{equation}
	\mu^{2}+(k-1)\,\mu-\omega^{2}=0.            \label{eqn:6.2.7}
	\end{equation}
In view of non-negativity of $\omega^2$ (the eigenvalue of
$-c^{ab}\nabla_{a}
\nabla_{b}$) these roots are of opposite signs, $\mu_{-}\mu_{+}=
-\omega^2\leq0$, and we can choose $\mu_{-}$ to be non-negative in
order to
have ${\mbox{\boldmath $u$}}_{-}(\tau)$ as a set of regular basis
functions at
$\tau_{-}=0$, the remaining part of them ${\mbox{\boldmath
$u$}}_{+}(\tau)$
beeing singular. By our assumption that the operator
${\mbox{\boldmath $F$}}$
does not have zero eigenvalues on the Euclidean spacetime of the
no-boundary
type, one can be sure that there are no basis functions which are
simultaneously regular at $\tau_{-}=0$ and vanishing at $\tau_{+}$.
Therefore,
all the functions ${\mbox{\boldmath $u$}}_{+}(\tau)$ subject to
Dirichlet
boundary conditions at $\tau_{+}$ have for $\tau\rightarrow 0$ a
singular
asymptotic behaviour (\ref{eqn:6.2.6}) with $\mu_{+}<0$.

Thus we can use the functions (\ref{eqn:6.2.5}) and (\ref{eqn:6.2.6})
satisfying respectively the regularity and the Dirichlet boundary
conditions as
the basis functions ${\mbox{\boldmath $u$}}_{\pm}$ of the previous
section,
construct by the same algorithm (\ref{eqn:6.17a}) the corresponding
Green's
function on spacetime of the no-boundary type and develope along the
same lines
the variational technique for the functional determinant. Thus we
arrive at the
same algorithm (\ref{eqn:6.28}), but due to the above properties of
the point
$\tau=\tau_{-}$ it can be further simplified. Indeed, in view of
(\ref{eqn:6.2.2}) the Wronskian operator takes the form
	\begin{equation}
	{\mbox{\boldmath
$W$}}(d/d\tau)=a_{0}\,\tau^{k}\,\frac{d}{d\tau}+
	O\,(\,\tau^{k+1}).
\label{eqn:6.2.8}
	\end{equation}
Therefore, the Wronskian matrix ${\mbox{\boldmath $\Delta$}}_{-+}$
calculated
at $\tau_{-}=0$ equals
	\begin{equation}
	{\mbox{\boldmath $\Delta$}}_{-+}=(\mu_{+}-\mu_{-})\;
	{\mbox{\boldmath $U$}}^{T}_{\!-}a_{0}{\mbox{\boldmath
$V$}}_{\!+},
\label{eqn:6.2.9}
	\end{equation}
because of a simple property $\mu_{-}\!+\mu_{+}=1-k$ of the roots of
eq.(\ref{eqn:6.2.7}). Using this result in the matrix $I_{+-}$
defined by
eq.(\ref{eqn:6.27}) together with the asymptotic behaviour
(\ref{eqn:6.2.5}) -
(\ref{eqn:6.2.6}) we get
	\begin{equation}
	\left(\,I_{+-}^{\,T}\right)^{-1}=
	{\mbox{\boldmath $u$}}_{-}(\tau_{+})\,
	({\mbox{\boldmath $\Delta$}}_{-+}^{T}\!)^{-1}{\mbox{\boldmath
$u$}}_{+}^{T}(\tau_{-})=
	{\rm Const}\,{\mbox{\boldmath $u$}}_{-}(\tau_{+})\,
	{\mbox{\boldmath $U$}}_{-}^{-1}.
\label{eqn:6.2.10}
	\end{equation}
Here we have absorbed all field-independent matrix multiplyers into
the overall
coefficient Const. This multiplyers include the infinite factor
$\tau^{\mu_{+}},\,\tau\rightarrow\tau_{-}=0$, the roots $\mu_{\pm}$
of the
equation (\ref{eqn:6.2.7}) and the configuration space metric $a_{0}$
which are
completely field independent, because they are defined with respect
to a
distingushed geometry of a 3-dimensional sphere of unit radius
embedded into
flat spacetime with unit radial lapse (in particular, the eigenvalues
$\omega^{2}$ entering (\ref{eqn:6.2.7}) belong to the covariant
Laplacian in
this distinguished metric $c_{ab}$)
\footnote
{The variational technique of the previous section gives the
functional
determinant of the differential operator only up to the overall
coefficient
independent of all the background field variables - the coefficients
of
${\mbox{\boldmath $F$}}$. This justifies the above procedure of
absorbing such
field-independent quantities into the overall normalization which
should be
defined from other principles, just like the normalization
coefficient in the
Pauli-Van Vleck-Morette formula follows from the composition law for
the kernel
(\ref{eqn:6.31}).
}.

Note that the basis functions are generally defined only up to their
linear
recombinations ${\mbox{\boldmath
$u$}}^{i}_{\pm\,A}(\tau)\rightarrow{\mbox{\boldmath
$u$}}^{i}_{\pm\,B}
(\tau)\,\Omega^{B}_{A}$ with certain time-independent coefficients
$\Omega^{B}_{A}$ which can be arbitrary functionals of fields
entering the
operator ${\mbox{\boldmath $F$}}$. In particular, the leading
coefficients
${\mbox{\boldmath $U$}}_{\!-}$ in (\ref{eqn:6.2.5}) may be such
functionals.
Thus, the right-hand side of (\ref{eqn:6.2.10}) can be viewed as a
new set of
regular basis functions having such a special normalization that
their leading
term in a small $\tau$ expansion is field-independent and is
proportional to
the functional matrix unity ${\mbox{\boldmath $I$}}$:
${\mbox{\boldmath
$u$}}_{-}(\tau)\,{\mbox{\boldmath $U$}}_{\!-}^{-1}={\mbox{\boldmath
$I$}}\,
\tau^{\mu_{-}}+ O\,(\,\tau^{1+\mu_{-}})$, $\tau\rightarrow 0$. The
nature of
the condensed index enumerating these new basis functions coincides
with that
of $i$: it is a covariant index $k$ originating in (\ref{eqn:6.2.10})
from
projecting the index $A$ of ${\mbox{\boldmath $u$}}^{i}_{-\,A}$ with
the matrix
$({\mbox{\boldmath $U$}}_{\!-}^{-1}) ^{\,A}_{\,k}$.

Thus we can write down the final algorithm for the one-loop prefactor
of the
no-boundary type together with the corresponding normalization of the
regular
basis functions in the center of the Euclidean ball
	\begin{eqnarray}
	&&\left(\,\frac{{\rm Det}\,{\mbox{\boldmath $F$}}}{{\rm
Det}\,
	{\mbox{\boldmath $a$}}}\,\right)^{-1/2}=
	{\rm Const}\,\left[\;{\rm det}\,
	{\mbox{\boldmath $u$}}_{-}(\tau_{+}\!)\;\right]^{-1/2},
\label{eqn:6.2.11}
	\end{eqnarray}
	\begin{eqnarray}
	\!\!\!\!\!\!{\mbox{\boldmath $u$}}_{-}(\tau)=
	{\mbox{\boldmath $I$}}\;\tau^{\mu_{-}}+
	O\,(\,\tau^{1+\mu_{-}}\!),\;\;\;
	\tau\rightarrow 0.                        \label{eqn:6.2.12}
	\end{eqnarray}

Apart from special terms responsible for the renormalization of
ultraviolet
infinities, which we don't consider here for reasons explained in
Sect.2, this
algorithm was previously derived in author's paper \cite{BKK}. This
derivation
was based on the technique of $\zeta$-functional regularization for
operators
with the explicitly unknown spectra on manifolds with a boundary (see
also
\cite{zeta,BKKM}).

\section{The functional determinant on closed spacetime without
boundary}
\hspace{\parindent}
Let us now consider the case of the closed compact Euclidean
spacetime without
boundary having a topology $S^{4}$. In analogy with the previous
section such a
spacetime can be obtained from the tube-like manifold
$[\tau_{-},\tau_{+}\!]
\times S^{3}$ by shrinking to the point both of its boundary surfaces
$S^{3}_{\pm}$ and imposing at these two points the no-boundary
regularity
conditions (Fig.3). Obviously, now the arising two poles $\tau_{\pm}$
of this
spherical manifold turn to be singular points of the radial part of
${\mbox{\boldmath $F$}}$ having similar asymptotic behaviours
(\ref{eqn:6.2.2})
of $a_{ik}$ (with $\tau$ replaced by $\tau-\tau_{-}$ and
$\tau_{+}-\tau$
respectively for $\tau\rightarrow\tau_{-}$ and
$\tau\rightarrow\tau_{+}$).
Therefore the two sets of basis functions ${\mbox{\boldmath
$u$}}_{-}(\tau)$
and ${\mbox{\boldmath $u$}}_{+}(\tau)$ regular respectively at the
south
$\tau=\tau_{-}$ and north $\tau=\tau_{+}$ poles of this spacetime
have the
following asymptotic behaviours
	\begin{eqnarray}
	{\mbox{\boldmath $u$}}_{-}(\tau)={\mbox{\boldmath
$U$}}_{\!-}\,
	(\tau-\tau_{-}\!)^{\mu_{-}}
	+O\,[\,(\tau-\tau_{-}\!)^{1+\mu_{-}}\!],
\label{eqn:6.3.1}\\
	{\mbox{\boldmath $u$}}_{+}(\tau)={\mbox{\boldmath
$V$}}_{\!+}\,
	(\tau-\tau_{-}\!)^{\mu_{+}}
	+O\,[\,(\tau-\tau_{-}\!)^{1+\mu_{+}}],
\label{eqn:6.3.2}
	\end{eqnarray}
for $\tau\rightarrow\tau_{-}$ and
	\begin{eqnarray}
	{\mbox{\boldmath $u$}}_{-}(\tau)={\mbox{\boldmath
$V$}}_{\!-}\,
	(\tau_{+}-\tau)^{\nu_{-}}
	+O\,[\,(\tau_{+}-\tau)^{1+\nu_{-}}\!],
\label{eqn:6.3.3}\\
	{\mbox{\boldmath $u$}}_{+}(\tau)={\mbox{\boldmath
$U$}}_{\!+}\,
	(\tau_{+}-\tau)^{\nu_{+}}
	+O\,[\,(\tau_{+}-\tau)^{1+\nu_{+}}],
\label{eqn:6.3.4}
	\end{eqnarray}
for $\tau\rightarrow\tau_{+}$. Here $\mu_{\pm}$ and $\nu_{\pm}$ are
pairs of
roots of the quadratic equations (\ref{eqn:6.2.7}) associated with
these two
poles. They are also completely field-independent and are chosen to
satisfy the
relations $\mu_{-}=\nu_{+}\geq 0$ and $\mu_{+}=\nu_{-}\leq 0$ in
accordance
with the regularity of ${\mbox{\boldmath $u$}}_{-}(\tau)$ at
$\tau_{-}$ and of
${\mbox{\boldmath $u$}}_{+} (\tau)$ at $\tau_{+}$ (note that we have
reserved
the notation ${\mbox{\boldmath $U$}}$ for the coefficients of regular
asymptotic behaviours in (\ref{eqn:6.3.1})- (\ref{eqn:6.3.4}) and the
notation
${\mbox{\boldmath $V$}}$ for those of singular ones).

Now, in addition to the expression (\ref{eqn:6.2.9}) for
${\mbox{\boldmath
$\Delta$}}_{-+}$ we can write down another expression for the same
quantity
calculated at $\tau_{+}$
	\begin{equation}
	{\mbox{\boldmath $\Delta$}}_{-+}=-(\nu_{+}-\nu_{-})\,
	{\mbox{\boldmath $V$}}^{T}_{\!-}a_{0}{\mbox{\boldmath
$U$}}_{\!+}
\label{eqn:6.3.5}
	\end{equation}
and get in virtue of (\ref{eqn:6.2.9}) the following relation for the
inverse
of the matrix $I_{+-}$
	\begin{equation}
	(\,I_{+-})^{-1}=
	{\rm Const}\,({\mbox{\boldmath $U$}}_{\!-}^{T})^{-1}
	{\mbox{\boldmath $V$}}_{-}^{T},
\label{eqn:6.3.6}
	\end{equation}
where we again absorb all field-independent matrix multiplyers into
the
coefficient Const. In view of (\ref{eqn:6.3.5}), however, this
relation can be
rewritten as
	\begin{equation}
	(\,I_{+-})^{-1}=
	{\rm Const}\,({\mbox{\boldmath $U$}}_{\!-}^{T})^{-1}
	{\mbox{\boldmath $\Delta$}}_{-+}\,({\mbox{\boldmath
$U$}}_{\!+})^{-1}
  \label{eqn:6.3.7}
	\end{equation}
and similarly to the previous section interpreted as a Wronskian
matrix
${\mbox{\boldmath $\Delta$}}_{-+}$ calculated according to its
definition
(\ref{eqn:6.12}) with respect to the specially normalized basis
functions
${\mbox{\boldmath $u$}}_{\pm}(\tau)\,{\mbox{\boldmath
$U$}}_{\!\pm}^{-1}$ which
have matrix-unitary behaviours of their regular asymptotics	at
$\tau_{\pm}$.

Therefore, the final algorithm for the one-loop prefactor on compact
spacetime
without boundary takes the form
	\begin{eqnarray}
	\left(\,\frac{{\rm Det}\,{\mbox{\boldmath $F$}}}{{\rm Det}\,
	{\mbox{\boldmath $a$}}}\,\right)^{-1/2}=
	{\rm Const}\,\left[\;{\rm det}\,
	{\mbox{\boldmath $\Delta$}}_{-+}\right]^{-1/2},
\label{eqn:6.3.8}
	\end{eqnarray}
where the Wronskian matrix is calculated with respect to the basis
functions
${\mbox{\boldmath $u$}}_{\pm}$ satisfying the following
field-independent
normalization at the points of their regular asymptotics
	\begin{eqnarray}
	{\mbox{\boldmath $u$}}_{-}(\tau)={\mbox{\boldmath $I$}}\,
	(\tau-\tau_{-}\!)^{\mu_{-}}\!
	+O\,[\,(\tau-\tau_{-}\!)^{1+\mu_{-}}\!],\,\,\,\,
	\tau\rightarrow\tau_{-},
\label{eqn:6.3.9}\\
	{\mbox{\boldmath $u$}}_{+}(\tau)={\mbox{\boldmath $I$}}\,
	(\tau_{+}-\tau)^{\nu_{+}}\!
	+O\,[\,(\tau_{+}-\tau)^{1+\nu_{+}}],\,\,\,\,
	\tau\rightarrow\tau_{+}.
\label{eqn:6.3.10}
        \end{eqnarray}

This algorithm has a deep analogy in the S-matrix theory in
asymptotically flat
and empty Lorentzian spacetime which admits the existence of the
so-called
standard IN and OUT vacua associated with the remote past and future
\cite{DW:Dynamical,DW:LesH}. These vacua are defined relative to the
positive-negative frequency decomposition of linearized quantum
fields with
respect to the basis functions ${\mbox{\boldmath
$v$}}_{IN}^{\phantom\dagger}$
and ${\mbox{\boldmath $v$}}_{OUT}^{\phantom\dagger}$ of the
Lorentzian wave
equation
	\begin{equation}
	{\mbox{\boldmath $F$}}_{\!L}(d/dt)\,
	{\mbox{\boldmath $v$}}_{IN}^{\phantom\dagger}(t)=0,\;\;\;
	{\mbox{\boldmath $F$}}_{\!L}(d/dt)\,
	{\mbox{\boldmath $v$}}_{OUT}^{\phantom\dagger}(t)=0,
\label{eqn:6.3.11a}
	\end{equation}
their complex conjugates (negative energy modes) having the following
behaviour
respectively at $t\rightarrow -\infty$ and $t\rightarrow +\infty$
	\begin{eqnarray}
	{\mbox{\boldmath $v$}}_{IN}^*\rightarrow
	{e}^{i\omega\,({\bf k}\!)\,t+
        i{\bf k}{\bf x}},\,\,\,\,t\rightarrow -\infty; \,\,\,\,\,\,
	{\mbox{\boldmath $v$}}_{OUT}^*\rightarrow
	{e}^{i\omega\,({\bf k}\!)\,t+
        i{\bf k}{\bf x}},\,\,\,\,t\rightarrow +\infty.
\label{eqn:6.3.11}
	\end{eqnarray}

As it was shown by DeWitt \cite{DW:Dynamical,DW:LesH}, the one-loop
transition
amplitude between these vacua in the presence of external sources,
reflecting
nontrivial field and geometry configuration in the interior of
spacetime, can
be represented in terms of the (functional) matrix $\alpha$ of
Bogolyubov
coefficients relating the IN and OUT sets of the basis functions
	\begin{eqnarray}
	{\mbox{\boldmath $v$}}_{OUT}^{\phantom\dagger}=
	\alpha\,{\mbox{\boldmath $v$}}_{IN}^{\phantom\dagger}+
	\beta\,{\mbox{\boldmath $v$}}_{IN}^*.
\label{eqn:6.3.12}
	\end{eqnarray}
Here $\alpha$ is calculable as a set of matrix elements between
${\mbox{\boldmath $v$}}_{IN}^{\phantom\dagger}$ and ${\mbox{\boldmath
$v$}}_{OUT}^{\phantom\dagger}$ with respect to the (indefinite) inner
product
in the space of solutions of the Lorentzian wave equation (relative
to which
the positive-energy basis functions are orthonormal)
	\begin{eqnarray}
	\alpha=<{\mbox{\boldmath $v$}}_{IN}^{\phantom\dagger},
	\,{\mbox{\boldmath $v$}}_{OUT}^{\phantom\dagger}>\equiv
	i\left[\,{\mbox{\boldmath $v$}}^{\dagger}_{IN}\,
	({\mbox{\boldmath
$W$}}_{\!\!L}^{\phantom\dagger}\,{\mbox{\boldmath
$v$}}_{OUT}
	^{\phantom\dagger}\!)-
	({\mbox{\boldmath $W$}}_{\!\!L}^{\phantom\dagger}\,
	{\mbox{\boldmath $v$}}_{IN}^{\dagger}\!)\,
	{\mbox{\boldmath $v$}}_{OUT}^{\phantom\dagger}\,\right],
\label{eqn:6.3.13}
	\end{eqnarray}
where ${\mbox{\boldmath $W$}}_{L}$ is a corresponding {\it
Lorentzian}
Wronskian operator of ${\mbox{\boldmath $F$}}_{\!L}$ and
${\mbox{\boldmath
$v$}}_{IN}^{\dagger}\equiv({\mbox{\boldmath $v$}}_{IN}^*)^{T}$.

In terms of $\alpha$ the one-loop contribution to the IN-OUT
transition
amplitude
	\begin{eqnarray}
	<OUT\,|\,IN>= {\rm e}^{\!\phantom {0}^{\textstyle
	\frac{i}{\hbar}S+iW_{\rm one-loop}}}
	+O\,(\,\hbar\,)
\label{eqn:6.3.14}
	\end{eqnarray}
looks simply as \cite{DW:Dynamical,DW:LesH}
	\begin{eqnarray}
	{\rm e}^{\!\phantom {0}^{\textstyle iW_{\rm one-loop}}}=
	{\rm Const}\,\left[\;{\rm det}\,
	\alpha\,\right]^{-1/2},
\label{eqn:6.3.15}
	\end{eqnarray}
which is just the algorithm (\ref{eqn:6.3.8}) with the Wronskian
matrix
${\mbox{\boldmath $\Delta$}}_{-+}$ replaced by the {\it Lorentzian}
Wronskian
matrix (\ref{eqn:6.3.13}) of the Bogolyubov coefficients. On the
other hand,
this one-loop factor also has a representation
	\begin{eqnarray}
	{\rm e}^{\!\phantom {0}^
	{\textstyle iW_{\rm one-loop}}}=
	\left(\frac{{\rm Det}\,{\mbox{\boldmath $F$}}_{\!L}}{{\rm
Det}\,
	{\mbox{\boldmath $a$}}}\right)^{-1/2}
\label{eqn:6.3.16}
	\end{eqnarray}
of the functional determinant of the Lorentzian wave operator
${\mbox{\boldmath
$F$}}_{\!L}$ (and the local measure) corresponding to the classical
action $S$
in eq.(\ref{eqn:6.3.14}), calculated on the space of functions having
asymptotic behaviour (\ref{eqn:6.3.11}). Thus, our algorithm
(\ref{eqn:6.3.8})
can be regarded as a direct {\it Euclidean closed space} analogue of
the
reduction method for the functional determinant of the hyperbolic
wave operator
in the {\it Lorentzian asymptotically flat} spacetime. Note that in
this
comparison the asymptotic behaviour (\ref{eqn:6.3.11}) at
$t\rightarrow\pm\infty$ plays the role of regularity conditions
(\ref{eqn:6.3.9}) - (\ref{eqn:6.3.10}) at smooth poles $\tau_{\pm}$
of the
Euclidean spherical manifold. This analogy becomes even deeper when
one
considers the analytic continuation of the theory in the Lorentzian
asymptotically flat spacetime to the Euclidean one with zero boundary
conditions at the asymptotically flat infinity -- the standard
calculational
method of Wick rotation \cite{CPTI}. Under this rotation, $\tau=it$,
the basis
functions (\ref{eqn:6.3.11}) go over into the basis functions of the
Euclidean
Klein-Gordon equation ${\mbox{\boldmath $u$}}^{\rm af} _{-}(\tau)\sim
{\rm
e}^{\omega({\bf k})\tau}$, $\tau\rightarrow -\infty$, and
${\mbox{\boldmath
$u$}}^{\rm af}_{+}(\tau)\sim {\rm e}^{-\omega({\bf k})\tau}$,
$\tau\rightarrow
+\infty$, satisfying zero boundary conditions respectively at
$\tau\rightarrow
-\infty$ and $\tau\rightarrow +\infty$, if we assume the following
analytic
continuation rule
	\begin{eqnarray}
	{\mbox{\boldmath $u$}}^{\rm af}_{-}(it)={\mbox{\boldmath
$v$}}_{IN}^*(t),\,\,\,\,
	{\mbox{\boldmath $u$}}^{\rm af}_{+}(it)={\mbox{\boldmath
$v$}}_{OUT}^
	{\phantom\dagger}(t).                      \label{eqn:6.3.17}
	\end{eqnarray}

Thus the functions ${\mbox{\boldmath $u$}}^{\rm af}_{\pm}(\tau)$,
vanishing at
$\tau\rightarrow\pm\infty$ and singular at the opposite asymptotics
$\tau\rightarrow\mp\infty$ serve as a direct asymptotically-flat
version of the
two sets of basis functions ${\mbox{\boldmath $u$}}_{\pm}(\tau)$
regular
respectively at the north $\tau_{+}$ and south $\tau_{-}$ poles of
our compact
manifold of spherical topology. By the analytic continuation rule
(\ref{eqn:6.3.17}) they give rise to two pairs of complex conjugated
basis
functions $({\mbox{\boldmath
$v$}}_{IN}^{\phantom\dagger},\,{\mbox{\boldmath
$v$}}_{IN}^*)$ and $({\mbox{\boldmath
$v$}}_{OUT}^{\phantom\dagger},\,{\mbox{\boldmath $v$}}_{OUT}^*)$
which
determine the standard vacua in Lorentzian quantum field theory, thus
relating
the notion of regularity and Dirichlet boundary conditions in
Euclidean
spacetime to the choice of preferred quantum states in the Lorentzian
one. As
was shown in \cite{tunnelI}, this formal analogy takes in the context
of
quantum cosmology the form of the physical process of nucleating the
Lorentzian
Universe with a special vacuum state from the Euclidean spacetime of
the
no-boundary type \cite{vacuum,Gibbons-Pohle}. The closed Euclidean
spacetime
also naturally arises in the calculation of the quantum distribution
of such
Universes as a gravitational quasi-DeSitter instanton carrying the
effective
action of the theory \cite{tunnelI}. Such a distribution weighted by
this
Euclidean action can be viewed as an analogue of the IN-OUT matrix
element
(\ref{eqn:6.3.14}) weighted by the Lorentzian effective action
${\mbox{\boldmath $\Gamma$}}_{\rm 1-loop}=S+\hbar\,W_{\rm 1-loop}$.
This
emphasizes a deep interrelation between the Euclidean and Lorentzian
quantum
theories, well-known in the S-matrix context as a Wick rotation
\cite{CPTI} and
extending it to quantum gravity of tunnelling geometries on spatially
closed
spacetimes \cite{tunnelI}.

\section*{Acknowledgements}
\hspace{\parindent}
The author benefitted from helpful discussions with G.Gibbons and is
also
grateful to A.Umnikov for his help in the preparation of the pictures
for this
paper. This work was supported by CITA National Fellowship and NSERC
grant at
the University of Alberta.

\appendix
\renewcommand{\thesection}{}
\section{Appendix. The basis function representation of the Green's
function}
\renewcommand{\theequation}{A.\arabic{equation}}
\hspace{\parindent}
Consider two complete sets of basis functions ${\mbox{\boldmath
$u$}}_{\pm}$ of
the operator ${\mbox{\boldmath $F$}}$ subject to boundary conditions
(\ref{eqn:6.10}). In virtue of these boundary conditions the
$\tau$-independent
matrix of their Wronskian inner products
$\varphi^{T}_1({\mbox{\boldmath
$W$}}\!\varphi_2)-({\mbox{\boldmath $W$}}\!\varphi_1)^{T}\varphi_2$,
$\varphi_{1,2}={\mbox{\boldmath $u$}}_{\pm}(\tau)$, can be rewritten
in the
form
      \begin{equation}
      \left[\begin{array}{cc}
      {\mbox{\boldmath $u$}}^{T}_{-}&-\,({\mbox{\boldmath
$W$}}\!{\mbox{\boldmath $u$}}_{-})^{T}\\
      {\mbox{\boldmath $u$}}^{T}_{+}&-\,({\mbox{\boldmath
$W$}}\!{\mbox{\boldmath $u$}}_{+})^{T}
      \end{array}\right]
      \left[\begin{array}{cc}
      {\mbox{\boldmath $W$}}\!{\mbox{\boldmath
$u$}}_{+}&{\mbox{\boldmath
$W$}}\!{\mbox{\boldmath $u$}}_{-}\\
      {\mbox{\boldmath $u$}}_{+}& {\mbox{\boldmath
$u$}}_{-}\end{array}\right]=
      \left[\begin{array}{cc}
      {\mbox{\boldmath $\Delta$}}_{-+}&0\\
      0&{\mbox{\boldmath $\Delta$}}_{+-}\end{array}\right],
\label{eqn:7.25}
      \end{equation}
where ${\mbox{\boldmath $\Delta$}}_{+-}$ and ${\mbox{\boldmath
$\Delta$}}_{-+}$
are mutually transposed matrices (\ref{eqn:6.11}) and
(\ref{eqn:6.12}). By
taking the determinant of this equation one immeadiately obtains, due
to the
factorized nature of its left hand side, the following relation
      \begin{equation}
      \left({\rm det}\left[\begin{array}{cc}
      {\mbox{\boldmath $W$}}\!{\mbox{\boldmath
$u$}}_{+}&{\mbox{\boldmath
$W$}}\!{\mbox{\boldmath $u$}}_{-}\\
      {\mbox{\boldmath $u$}}_{+}& {\mbox{\boldmath $u$}}_{-}
      \end{array}\right]\,\right)^{2}=
      (\,{\rm det}\,{\mbox{\boldmath $\Delta$}}_{+-}\!)^{2}.
\label{eqn:7.26}
      \end{equation}
This relation implies that the matrix ${\mbox{\boldmath
$\Delta$}}_{+-}$ can be
degenerate only when there is a linear dependence between
${\mbox{\boldmath
$u$}}_{+}$ and ${\mbox{\boldmath $u$}}_{-}$, which is ruled out by
our
assumption that the operator ${\mbox{\boldmath $F$}}$ does not have
zero modes
and is uniquely invertible under the Dirichlet boundary conditions at
$\tau_{\pm}$. Calculating this constant matrix at $\tau_{\pm}$ shows
also that
${\mbox{\boldmath $u$}}_{-}(\tau_{+})$ and ${\mbox{\boldmath
$u$}}_{+}(\tau_{-})$ are guaranteed to be invertible.

The equation (\ref{eqn:7.25}) has a simple corollary relying on the
proved
invertibility of matrices (\ref{eqn:6.11}) and (\ref{eqn:6.12})
      \begin{equation}
      \left[\begin{array}{cc}
      {\mbox{\boldmath $W$}}\!{\mbox{\boldmath
$u$}}_{+}&{\mbox{\boldmath
$W$}}\!{\mbox{\boldmath $u$}}_{-}\\
      {\mbox{\boldmath $u$}}_{+}& {\mbox{\boldmath
$u$}}_{-}\end{array}\right]^{-1}=
      \left[\begin{array}{cc}
      ({\mbox{\boldmath $\Delta$}}_{-+}\!)^{-1}&0\\
      0& ({\mbox{\boldmath $\Delta$}}_{+-}\!)^{-1}\end{array}\right]
      \left[\begin{array}{cc}
      {\mbox{\boldmath $u$}}^{T}_{-}&-\,({\mbox{\boldmath
$W$}}\!{\mbox{\boldmath $u$}}_{-})^{T}\\
      {\mbox{\boldmath $u$}}^{T}_{+}&-\,
      ({\mbox{\boldmath $W$}}\!{\mbox{\boldmath
$u$}}_{+})^{T}\end{array}\right].      \label{eqn:7.27}
      \end{equation}
When multipled from the left by the matrix entering the left hand
side of
(\ref{eqn:7.26}), this equation gives, on account of the form of the
Wronskian
operator (\ref{eqn:6.5a}), the relations (\ref{eqn:6.15}) and
(\ref{eqn:6.16})
for equal-time bilinear combinations of basis functions.

To derive the needed basis function representation (\ref{eqn:6.17a})
of the
Green's function of ${\mbox{\boldmath $F$}}$ subject to Dirichlet
boundary
conditions, let us consider the boundary value problem
	\begin{equation}
	{\mbox{\boldmath
$F$}}(d/d\tau)\,\varphi\,(\tau)=J\,(\tau),\;\;\;
	\varphi\,(\tau_{\pm})=0
\label{eqn:7.28}
	\end{equation}
with arbitrary source $J\,(\tau)$ for the function $\varphi\,(\tau)$
which can
be decomposed in the sets of basis functions ${\mbox{\boldmath
$u$}}_{\pm}(\tau)$ with some unknown time-dependent coefficients
$\varphi_{\pm}(\tau)$ satisfying the necessary boundary conditions:
	\begin{equation}
	\varphi\,(\tau)={\mbox{\boldmath $u$}}_{+}(\tau)\,
	\varphi_{+}(\tau)+
	{\mbox{\boldmath $u$}}_{-}(\tau)\,\varphi_{-}(\tau),\;\;\;
	\varphi_{\pm}(\tau_{\mp})=0.
\label{eqn:7.29}
	\end{equation}
Substituting this representation into the left hand side of the
equation
(\ref{eqn:7.28}), one finds that it can be satisfied if these
coefficients
$\varphi_{\pm}(\tau)$ solve the following system of equations
	\begin{equation}
	\left[\begin{array}{cc}
        {\mbox{\boldmath $W$}}\!{\mbox{\boldmath
$u$}}_{+}&{\mbox{\boldmath
$W$}}\!{\mbox{\boldmath $u$}}_{-}\\
        {\mbox{\boldmath $u$}}_{+}& {\mbox{\boldmath
$u$}}_{-}\end{array}\right]
	\left[\begin{array}{c}
        \dot\varphi_{+}\\
        \dot\varphi_{-}\end{array}\right]=
        -\left[\begin{array}{c}J\\0\end{array}\right],
	 \;\;\;\;\;\;\dot\varphi\equiv d\varphi/d\tau.
\label{eqn:7.30}
	 \end{equation}
In view of (\ref{eqn:7.27}) this linear system can be easily solved
with
respect to $\dot\varphi_{\pm}$ and integrated subject to boundary
conditions in
(\ref{eqn:7.29}), whence this decomposition takes the form of the
integral
	\begin{equation}
	\varphi\,(\tau)=\int_{\tau_{-}}^{\tau_{+}}d\tau'\,
	{\mbox{\boldmath $G$}}\,(\tau,\tau')\,J\,(\tau')
	\end{equation}
with the needed Green's function ${\mbox{\boldmath
$G$}}\,(\tau,\tau')$ given
by the equation (\ref{eqn:6.17a}).

\newpage
\section*{\centerline{Figure captions}}
{\bf Fig.1} The Euclidean spatially closed spacetime interpolating
between two
regular hypersurfaces $\Sigma_{\pm}$ of constant Euclidean time
$\tau_{\pm}$,
which underlies the heat-equation transition amplitude between the
configurations on these hypersurfaces.
\\
\\
{\bf Fig.2}  Euclidean spacetime of the no-boundary type originating
from the
tube-like manifold $\Sigma\times[\,\tau_{-},\tau_{+}]$ by shrinking
one of its
boundaries $\Sigma_{-}$ to a point which becomes a regular internal
point
$\tau_{-}$ of the resulting 4-dimensional ball.
\\
\\
{\bf Fig.3} Obtainig the closed compact spacetime of spherical
topology as a
continuation of the process depicted on Fig.2: shrinking the
remaining boundary
$\Sigma_{+}$ to a point $\tau_{+}$ and imposing at this point the
regularity
conditions of the no-boundary type. The resulting manifold inherits
the
foliation with slices of constant Euclidean time $\tau$ in the form
of its
quasi-spherical latitudinal sections $\Sigma$.

\end{document}